\documentclass{aa} 

\usepackage{graphics} 

\begin{document}

   \thesaurus{06 
              (08.05.2;
		11.13.1;
		11.19.4)} 
   \title{Be stars in and around young clusters in the Magellanic Clouds}

   \author{S. C. Keller
	  \and
	  P. R. Wood
          \and
          M. S. Bessell
          }

   \offprints{S. C. Keller}

   \institute{Mount Stromlo \& Siding Spring Observatories, Private Bag, Weston
    Creek PO, ACT 2611, Australia\\
              email: stefan@mso.anu.edu.au, wood@mso.anu.edu.au,
                     bessell@mso.anu.edu.au }

   \date{Received ; accepted}

   \titlerunning{Be stars in the Magellanic Clouds}


   \maketitle

\begin{abstract} 

We present the results of a search for Be stars in six fields centered on the
young clusters NGC 330 and NGC 346 in the SMC, and NGC 1818, NGC 1948, NGC 2004
and NGC 2100 in the LMC.  Be stars were identified by differencing R band and
narrow-band H$\alpha$ CCD images.  Our comparatively large images provide
substantial Be star populations both within the clusters and in their
surrounding fields. Magnitudes, positions and finding charts are given for the
224 Be stars found.  The fraction of Be stars to normal B stars within each
cluster is found to vary significantly although the average ratio is similar to
the average Be to B star ratio found in the Galaxy.  In some clusters, the Be
star population is weighted to magnitudes near the main sequence turn-off.  The
Be stars are redder in $V$$-$$I$ than normal main-sequence stars of similar
magnitude and the redness increases with increasing H$\alpha$ emission
strength.


\end{abstract}

\section{Introduction}

In this paper we present the results of a photometric survey for Be stars
within and around six young clusters in the Magellanic Clouds.  Such studies
have been carried out on Galactic clusters but they are limited there by the
relative paucity of Be stars. The young populous Magellanic Cloud clusters
examined in this paper present us with large samples of Be stars in a range of
metallicity environments and the opportunity to further examine the mechanism
for formation of Be stars.

Searches for Be stars in the Magellanic Clouds have been made by Feast (\cite{fea72})
using objective prism techniques, and the more recently by Grebel et
al. (\cite{gre92}), Bessell and Wood (\cite{bes92}) (who gave a brief description of
some of the results presented in this paper) and Grebel (\cite{gre97}) using
CCD-based imaging photometry. These latter searches involved imaging through an
H$\alpha$ filter so that Be stars, which show strong H$\alpha$ emission, stand
out in comparison to normal stars.  Photometric techniques such as these are very
efficient methods of identifying Be stars, in particular within dense
clusters where spectroscopy is difficult.

The main purpose of the present study is to provide a sample of Be stars for
follow-up spectroscopy and to see if there are differences in Be star
populations between different clusters and between clusters and the local field
population.  The results of the follow-up spectroscopy will be presented in a
subsequent paper.  Because the present study is ground-based, its spatial
resolution is seeing limited, which means that we are unable to search for Be
stars in the cluster cores.  In another paper, we will present the results of a
search for Be stars in the cluster cores using HST imaging.  The present paper
provides the comparison sample of Be stars in the local field and the outer
cluster zones.  The complete data set of
ground-based and HST images will allow an accurate comparison of field and
cluster Be stars properties.  In this paper, however, we make some initial
comparisons with the more limited data set described herein.

\section{Observations and Data Reduction}\label{obsred}

The photometric search method we used to detect Be stars involved taking a CCD
image through a narrow-band (15 \AA) H$\alpha$ filter and comparing this image
with a similar image obtained in the Cousins $R$ band.  Stars with strong
H$\alpha$ emission should appear brighter in the H$\alpha$ image than in the R
image.  Given that a typical Be star has H$\alpha$ emission with a full width
at half maximum of $\sim$7 \AA\ and peak H$\alpha$ flux 5 times the local
continuum flux, this sort of search with a 15 \AA\ filter is readily capable of
detecting Be stars.

The observations were obtained using direct CCD imaging on the 1-m telescope at
Siding Spring Observatory from September to November 1991.  All H$\alpha$
images were 900 second exposures. $R$ images were obtained immediately before
or after each H$\alpha$ image in order to avoid the possibility of stellar
variability affecting our measure of emission strength.  A Tektronix
1024$\times$1024 CCD with pixels of scale 0.6\arcsec\ was used for these
observations, giving us an imaged area of typically 
10\arcmin$\times$10\arcmin\ (compared with 
5.7\arcmin$\times$5.7\arcmin\ in Grebel \cite{gre97}).  
$V$ and $I$ images were obtained at the same time for some
fields, and in some cases on other nights using a GEC 770$\times$1152 CCD. A
log of the observations is presented in Table~\ref{obslog}.

The CCD images were processed with IRAF and the photometry of the fields was
done using the DoPHOT photometry package (Mateo and Schechter
\cite{mat89}). Standard magnitudes in $V$ and $I$ were derived from the
standards used in Sebo and Wood (\cite{seb94}) and from magnitudes given by Walker
(\cite{wal96}).  The $R$ and H$\alpha$ magnitudes were not standardised and
$R$$-$H$\alpha$ colours have an arbitrary zeropoint.

\begin{table} 
\caption{The log of observations presented in the current paper.}
\label{obslog} 
\begin{center} 
\begin{tabular}{cccc} 
\hline 
Cluster & Date(1991) & Filter & Exposure Time\\ 
\hline 
NGC 330 & Sept. 04 & V & 300s\\
        & & I & 150s	\\  
        & Oct. 31 & R & 300s\\	  
        & & H$\alpha$& 900s\\ 	  
NGC 346 & Nov. 29 & V & 150s\\	  
        & & I & 100s\\	  
        & & R & 150s\\  
        & & H$\alpha$& 900s\\
NGC 1818& Sept. 04 & V & 80s\\  
        & & I & 150s\\  
        & Oct. 31 & R & 300s\\	   
        & & H$\alpha$& 900s\\ 
NGC 1948& Oct. 30 & V & 120s\\
	& & I & 120s\\
	& & R & 200s\\
	& & H$\alpha$& 900s\\  
NGC 2004& Nov. 29 & V & 20s\\  
 	& & V & 150s\\           
  	& & I & 15s\\            
	& & I & 100s\\           
  	& & R & 15s\\            
 	& & R & 100s\\           
 	& & H$\alpha$ & 900s\\   
NGC 2100& Nov. 29 & V & 120s\\  
 	& & I & 100s\\           
 	& & R & 100s\\           
 	& & H$\alpha$ & 900s\\   

\hline
\end{tabular} 
\end{center} 
\end{table}

\section{Selection of Be and other H$\alpha$ emitting stars}\label{selection}

\begin{figure}
\resizebox{\hsize}{!}{\includegraphics{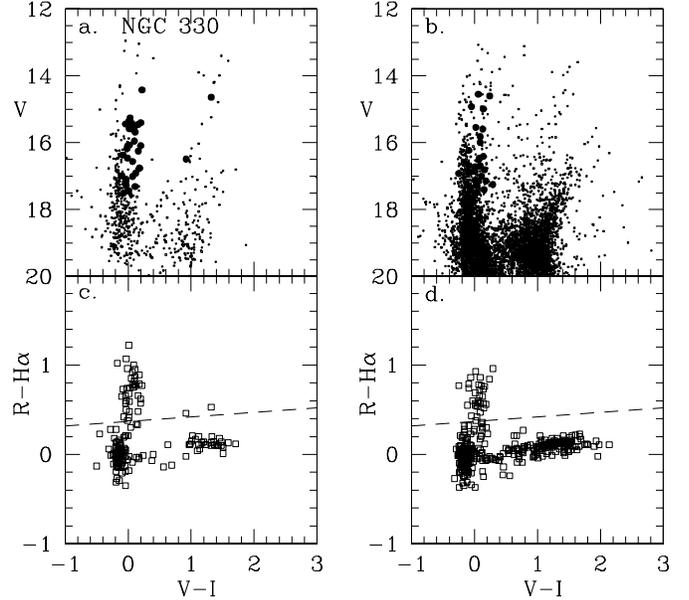}}
\caption{The ($V$, $V$$-$$I$) and ($R$$-$H$\alpha$, $V$$-$$I$) diagrams for the
cluster NGC 330 (panels {\bf a} and {\bf c}) and its surrounding field (panels
{\bf b} and {\bf d}). The usable cluster area lies between radii
$15\arcsec<r<1\arcmin.6$ while the field is defined as $r>1\arcmin.6$.  Be
stars are identified as the stars above the dashed line in the
($R$$-$H$\alpha$, $V$$-$$I$) diagrams and they are shown on the ($V$,
$V$$-$$I$) diagrams as filled circles.}
\label{fig330cmd} 
\end{figure}

\begin{figure}
\resizebox{\hsize}{!}{\includegraphics{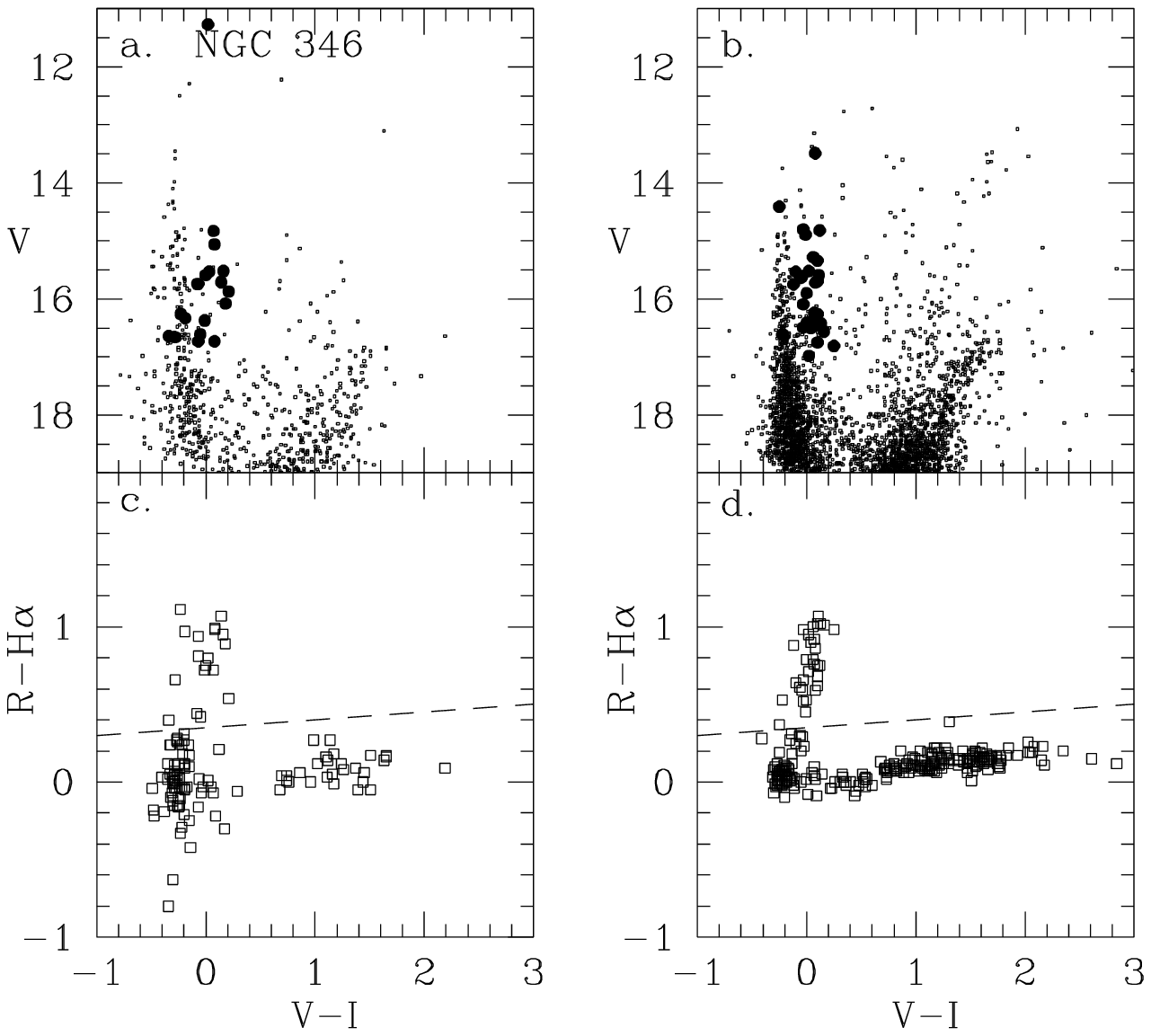}}
\caption{The same as Fig.~\ref{fig330cmd}, but for NGC 346. The usable cluster area
lies between radii $15\arcsec<r<2\arcmin.2$ while the field is
defined as $r>2\arcmin.2$.}
\label{fig346cmd} 
\end{figure}

\begin{figure}
\resizebox{\hsize}{!}{\includegraphics{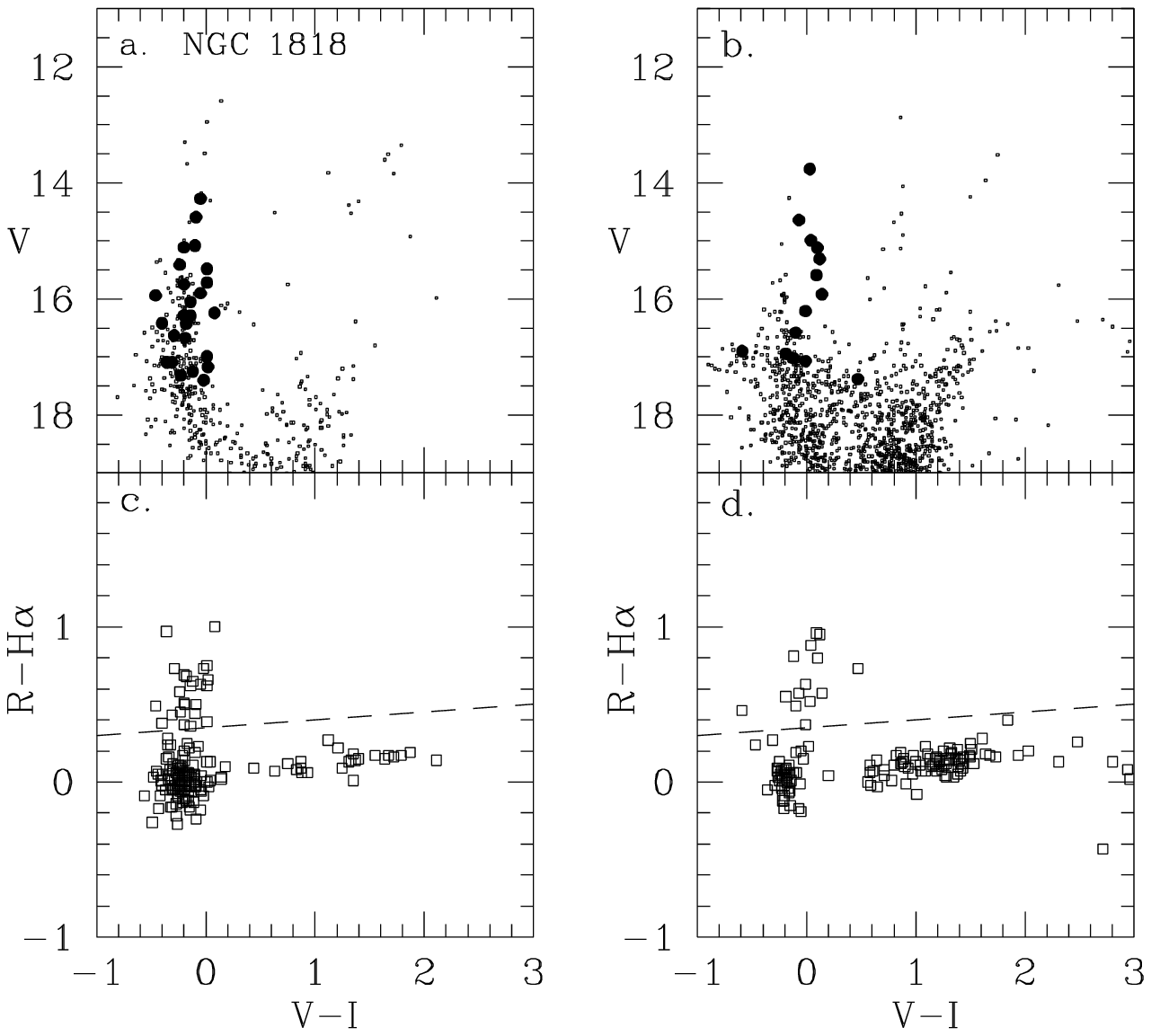}}
\caption{The same as Fig.~\ref{fig330cmd}, but for NGC 1818. The usable cluster area
lies between radii $15\arcsec<r<2\arcmin.2$ while the field is
defined as $r>2\arcmin.2$.}
\label{fig1818cmd} 
\end{figure}

\begin{figure}
\resizebox{\hsize}{!}{\includegraphics{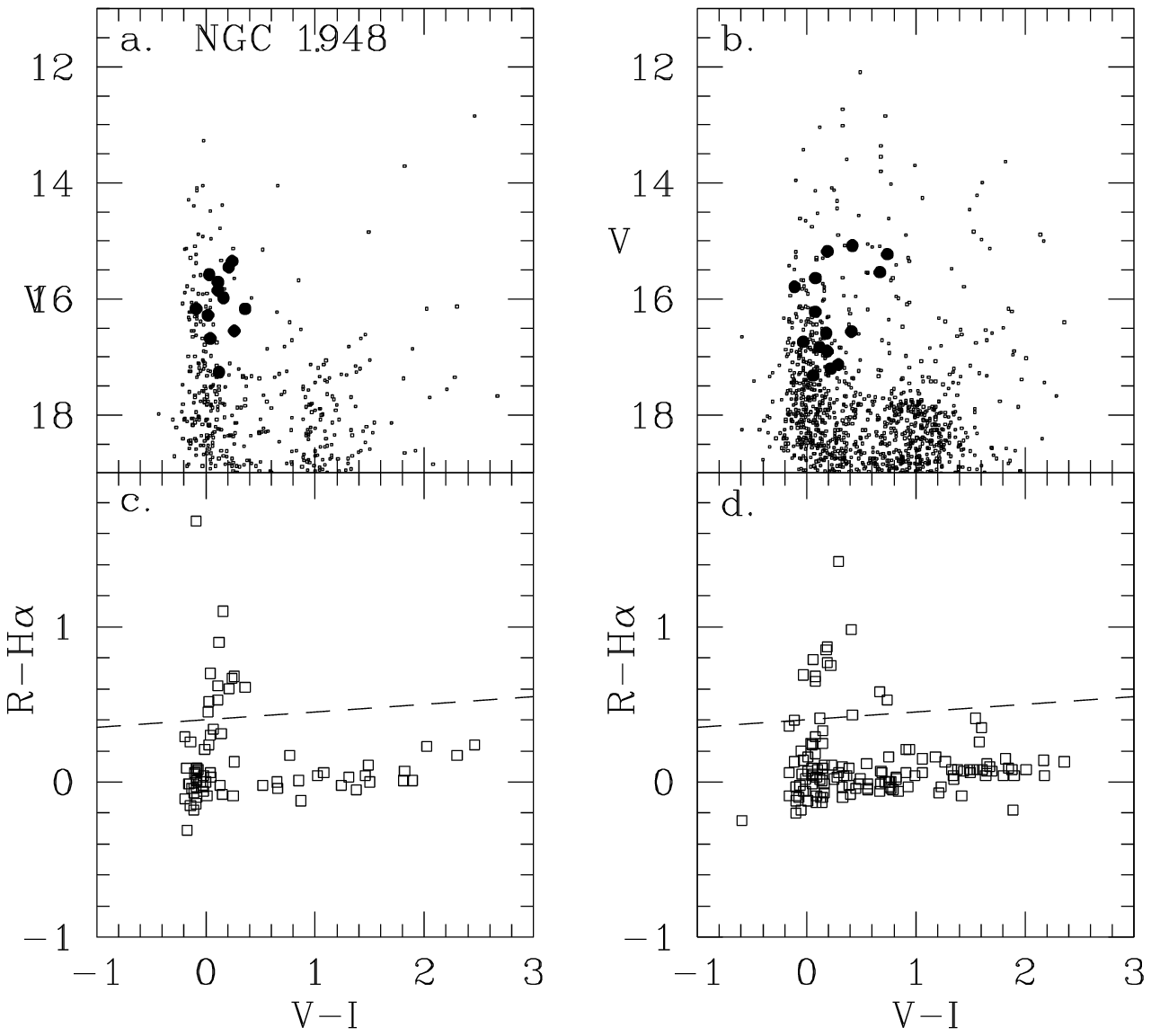}}
\caption{The same as Fig.~\ref{fig330cmd}, but for NGC 1948. The usable cluster area
lies between radii $15\arcsec<r<2\arcmin.7$ while the field is
defined as $r>2\arcmin.7$.}
\label{fig1948cmd} 
\end{figure}

\begin{figure}
\resizebox{\hsize}{!}{\includegraphics{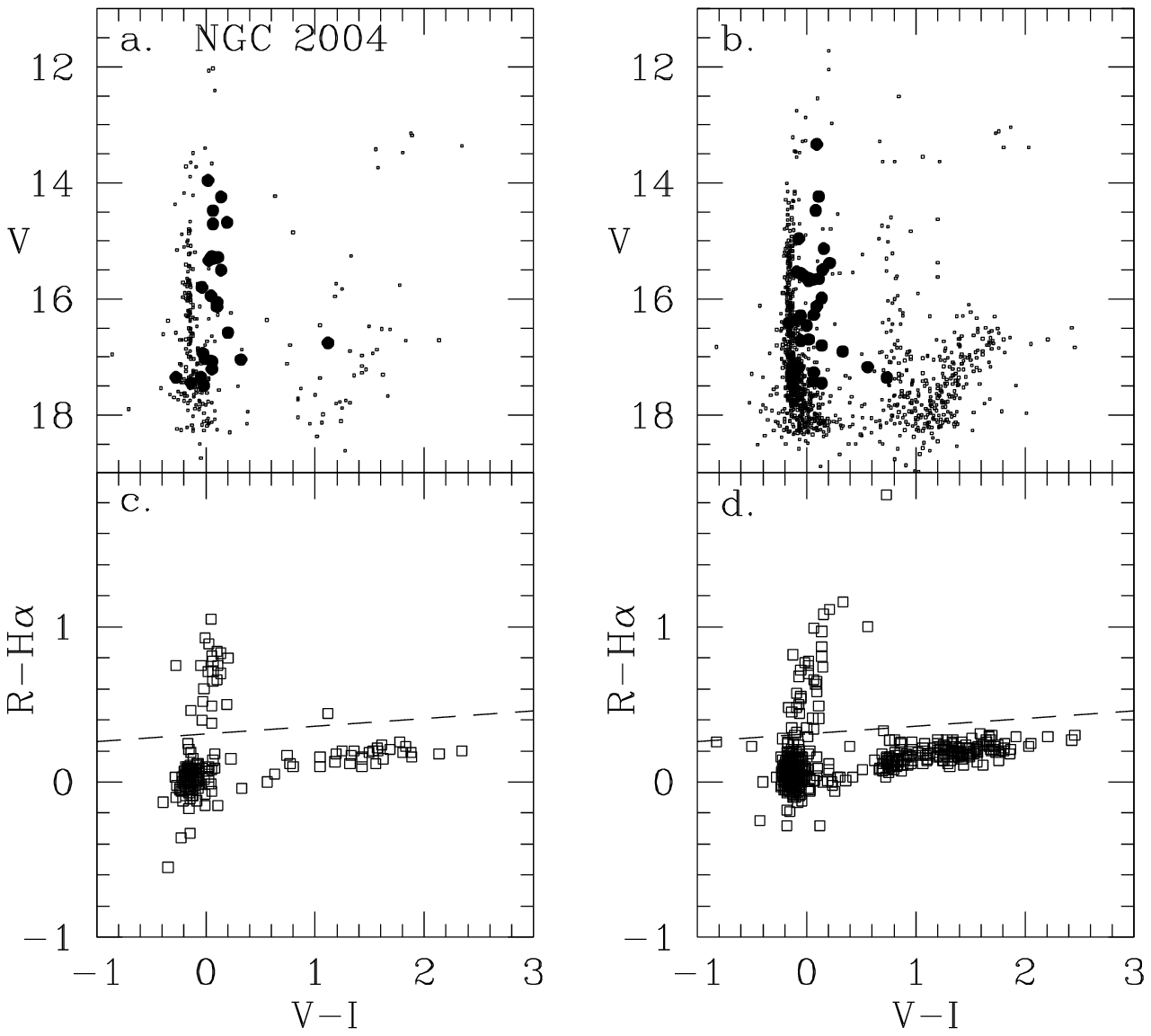}}
\caption{The same as Fig.~\ref{fig330cmd}, but for NGC 2004. The usable cluster area
lies between radii $15\arcsec<r<2\arcmin.0$ while the field is
defined as $r>2\arcmin.0$.}
\label{fig2004cmd} 
\end{figure}

\begin{figure}
\resizebox{\hsize}{!}{\includegraphics{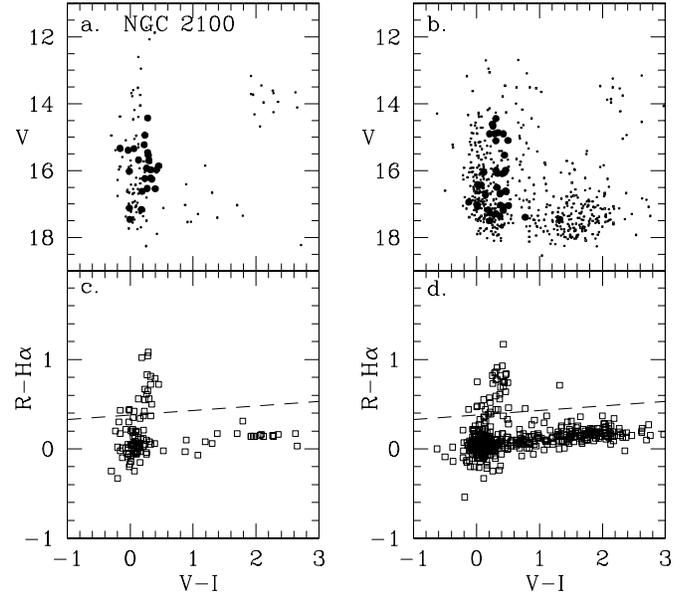}}
\caption{The same as Fig.~\ref{fig330cmd}, but for NGC 2100. The usable cluster area
lies between radii $15\arcsec<r<1\arcmin.4$ while the field is
defined as $r>1\arcmin.4$.}
\label{fig2100cmd} 
\end{figure}

In order to detect stars with H$\alpha$ emission, plots were made of the colour
index $R$$-$H$\alpha$ against $V$$-$$I$ for all the stars which were common to
the $V$, $I$, $R$ and H$\alpha$ frames. Figs.~\ref{fig330cmd} (NGC 330),
\ref{fig346cmd} (NGC 346), \ref{fig1818cmd} (NGC 1818), \ref{fig1948cmd} (NGC
1948), \ref{fig2004cmd} (NGC 2004) and \ref{fig2100cmd} (NGC 2100) show the
($R$$-$H$\alpha$, $V$$-$$I$) diagrams produced.  As noted in
Section~\ref{obsred}, the $R$$-$H$\alpha$ zero-point is arbitrary in these
plots, although the zero-point was chosen so that normal main-sequence stars
had $R$$-$H$\alpha$ $\approx$ 0.

The stars with strong H$\alpha$ emission were selected from the above figures.
Using Fig.~\ref{fig330cmd}c as a typical case, we note that most main-sequence
stars lie in a tight clump around $R$$-$H$\alpha$=0.0 and $V$$-$$I$=$-0.15$,
while nearly all cooler stars form an almost horizontal band. These two groups
of stars do not have detectable H$\alpha$ emission. However, there is another
group of stars extending up and slightly to the red of the main-sequence
clump: these stars clearly show significant H$\alpha$ emission and they are Be
stars.  Occasional cooler (giant) stars also exhibit strong H$\alpha$
emission.
  
The dispersion in $R$$-$H$\alpha$ of normal MS stars about the centre of the
clump is determined largely by the photometric errors in the
narrow-band H$\alpha$ magnitude, and this uncertainty will depend on the
magnitude of the star.  We limit our Be star selection process to stars with
$V$$<$17.5, which corresponds approximately to an error in $R$$-$H$\alpha$
colour of $\sim$0.2 magnitudes.  In order to select a sample of Be stars, we
draw a line in each of the $R$$-$H$\alpha$ diagrams parallel to the sequence of
non-emission line stars and at a distance of $\sim$0.4 magnitudes above it, which
gives a high probability that we exclude all members of the clump of normal
main-sequence stars.
 
Our observations were made under seeing conditions of 1.5--2\arcsec, and as a
consequence the central regions of the clusters suffer from crowding.  We
therefore exclude the innermost 15\arcsec\ (in radius) of each cluster from the
Be star search.  On each image taken, we select a radius at which the
stellar density drops to a value indistinguishable from that of the surrounding
field.  Stars within this radius are assumed to belong to the cluster, and
those outside to the field.  The radius selected for each cluster is given in
the captions of Figs.~\ref{fig330cmd} to \ref{fig2100cmd}.

The positions, magnitudes and colours of the Be stars selected as described
above are listed in Tables 2 (NGC 330), 3 (NGC 346),
4 (NGC1818), 5 (NGC 1948), 6 (NGC 2004) and 7 (NGC 2100). Tables 2-7 are available in electronic
form from the CDS. The astronomical coordinates in these tables were derived
from the secondary astrometric standards of Tucholke et
al. (\cite{tuc96}). Finding charts for the Be stars are given in Figs.~\ref{fig330field} and \ref{fig330cluster} (NGC 330), \ref{fig346field}
and \ref{fig346cluster} (NGC 346), \ref{fig1818field} and \ref{fig1818cluster}
(NGC 1818), \ref{fig1948field} (NGC 1948), \ref{fig2004field} and
\ref{fig2004cluster} (NGC 2004), and \ref{fig2100field} and
\ref{fig2100cluster} (NGC 2100).

\section{Comparison with past surveys for Be stars}

We now compare the results of our study to those of previous Be star surveys
within NGC 330 and NGC 1818 in order to get some idea of our detection
efficiency.  Although there have been a few reports of detections of Be stars
in other clusters (e.g. Kjeldsen and Baade \cite{kje94}), it is only for NGC
330 and NGC 1818 that any information is available regarding the identity of
the Be stars.

\subsection{NGC 330}
 
Grebel et al. (\cite{gre92}) have searched the region close to NGC 330 for Be
stars using narrow-band H$\alpha$ photometry.  Within 50\arcsec\ of the cluster
centre they find 29 Be stars while we find 32.  However, the Be star samples
found in the two respective searches are not completely identical.  Of the 18
Be stars for which Grebel et al. provide identifications (they list names from
Robertson \cite{rob74} and Arp \cite{arp59}), we find 16.  We suspect that
variability of the Be phenomenon may account for the differences in the two Be
star samples (see Section~\ref{n1818comp}).  Overall, however, the two searches
find similar numbers of stars. Our search area is considerably larger than that
of Grebel et al.  which will allow us to compare the properties of reasonable
samples of both field and cluster Be stars.

NGC 330 and its surroundings have been searched extensively for variable stars
by Sebo and Wood (\cite{seb94}) and Balona (\cite{bal92}).  A number of the Be
stars and one of the H$\alpha$ emitting red giants found here are known
variables.  The variable star identifications are listed in Table 2.  A
spectrum of the variable red giant was obtained by Sebo and Wood (\cite{seb94})
and it does indeed show a broad H$\alpha$ emission line.

\subsection{NGC 1818}\label{n1818comp}

Grebel (\cite{gre97}) presents the results of a Be star search within a 5.7\arcmin
$\times$ 5.7\arcmin\ field around NGC 1818.  Outside a radius of 15\arcsec\ from
the cluster centre, a total of 48 Be stars were found.  The study of Grebel
(\cite{gre97}) was made under excellent seeing conditions and has a limiting
magnitude of $V \sim$ 20.  Grebel's selection criterion required Be stars to be
$\sim$0.2 magnitudes above the main-sequence clump in the ($R$$-$H$\alpha$, $V$$-$$I$)
diagram, compared to our $\sim$0.35 magnitudes (see Grebel's Fig. 3 with our
Fig.~\ref{fig1818cmd}b).  However, given our H$\alpha$ filter bandpass of 15
\AA\ and Grebel's bandpass of 33 \AA, the sensitivities of the two surveys should be
similar, at least for the brighter stars where photon noise in the H$\alpha$
filter is $\la$0.35 magnitudes in our case, or 0.2 magnitudes in Grebel's case.

Grebel (\cite{gre97}) finds 21 Be stars with $V <$ 17.5 (the magnitude limit of
our study), of which we have identified 13 in common. Of the 8 unidentified
stars, 7 are near our magnitude limit and have 17.1 $< V <$ 17.5: furthermore,
most lie in crowded regions.  More interestingly, we detected 6 Be stars that
have no counterparts in the survey of Grebel.  This is perhaps surprising given
the fainter magnitude limit and better seeing of the Grebel survey. These
stars, which are all comparatively bright and lie in uncrowded regions, are
unlikely to be misidentifications.  We thus seem to have strong evidence for
the episodic nature of the Be phenomenon in Magellanic Cloud stars: such
behaviour is frequently seen in galactic Be stars (Hanuschik et
al. \cite{han93}).  It is also possible that some of the Be stars found by
Grebel, but which we failed to detect, were not exhibiting H$\alpha$ emission
at the time of our observation.

In summary, both the present survey and the surveys of Grebel et
al. (\cite{gre92}) and Grebel (\cite{gre97}) find similar numbers of Be stars,
although our survey may be incomplete near the magnitude limit of $V =$ 17.5
in crowded regions.  Because of the episodic nature of the Be phenomenon,
single-epoch surveys may miss a significant fraction of the total Be star
population ($\sim$20\% appears to have been missed in the case of NGC 1818).

\section{Photometric properties of the Be star population} 

There are a number of interesting correlations between the photometric indices
of Be stars observed here which may shed some light on the nature of the Be
star phenomenon.  Starting with Figs.~\ref{fig330cmd} to \ref{fig2100cmd}, we
see that within the clusters, as well as in the field, Be stars exist over a
wide range of luminosities.  Since the Be stars in the clusters are all of
essentially the same age (field contamination of the cluster Be star
populations should be $<$10\% based on the small cluster area), the wide range
in Be star luminosities means that the Be phenomenon can occur at any time
throughout the main-sequence lifetime, and need not be confined to particular
evolutionary phases such as the core contraction phase at the end of the
main-sequence.

Another interesting correlation is seen in the ($R$$-$H$\alpha$, $V$$-$$I$)
plots (Figs.~\ref{fig330cmd}c to \ref{fig2100cmd}c).  Here we see a significant
trend for the strongest emitters, as measured by the $R$$-$H$\alpha$ colour, to
be those with the reddest $V$$-$$I$.  This effect was also noted by Grebel
(\cite{gre97}).  The fact that Be stars are redder in $V$$-$$I$ than normal
B stars is also obvious from Figs.~\ref{fig330cmd}a,b to \ref{fig2100cmd}a,b
where, at a given $V$, the Be stars are clearly redder than the normal
main-sequence stars.  An increase in redness of Be stars with increased 
strength of H$\alpha$ emission has also been noted in the infrared (e.g. Dachs et
al. \cite{dac88}).  

It is widely accepted that the Be phenomenon is associated with rapid rotation
of the stellar photosphere and the presence of a circumstellar disk of
comparatively cool material which gives rise to the observed H$\alpha$
emission.  The observed increase in $V$$-$$I$ of Be stars with increased
strength of H$\alpha$ emission 
is almost certainly the result of one or more of four
factors: a real reduction in stellar effective temperature due perhaps to
rotation, a changed spectral energy distribution due to rotational distortion
of the stellar atmosphere, continuum and line emission from the disk, or
absorptive reddening caused by circumstellar matter. The first two factors
result from changes in the structure of the star, while the latter two involve
circumstellar matter only.  The viewing angle will also effect the colours in
the non-spherical geometry.

\begin{figure}
\resizebox{\hsize}{!}{\includegraphics{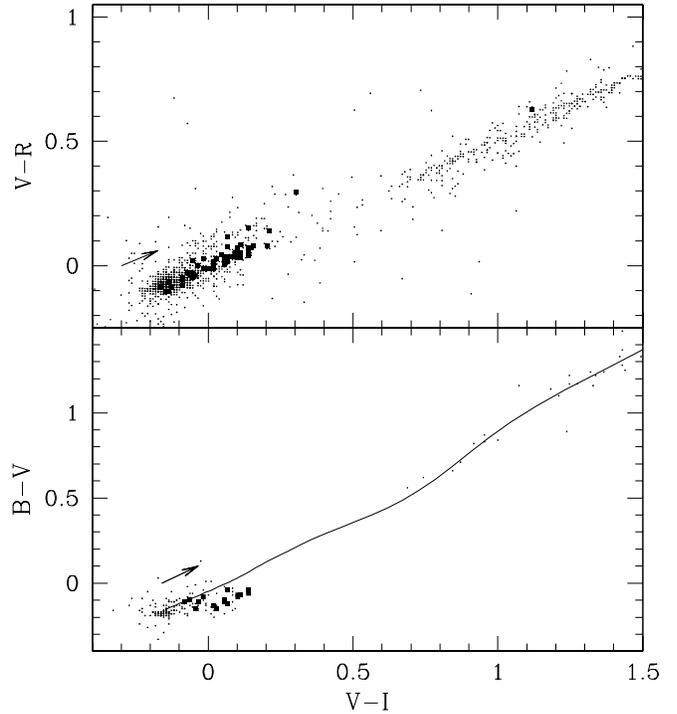}}
\caption{The ($V$$-$$R$, $V$$-$$I$) and ($B$$-$$V$, $V$$-$$I$) diagrams for
stars within and around the cluster NGC 2004, and brighter than
$V$=16.5. Non-emission B stars are shown as dots and Be stars are shown as
solid squares.  The $B$$-$$V$ colours are taken from Balona and Jerzykiewicz
(\cite{jerky}).  Due to the relatively small field studied by Balona and
Jerzykiewicz, there are fewer stars in the ($B$$-$$V$, $V$$-$$I$) diagram.  The
solid line in the ($B$$-$$V$, $V$$-$$I$) diagram is the theoretical
main-sequence ($\log g$ = 4.5) from Bessell et al. (\cite{bes98}).  The arrows
are interstellar reddening vectors from Taylor (\cite{tay86}), corresponding to
E($B$$-$$V$) = 0.1.}
\label{bmvvmiplot}
\end{figure}

In this regard our photometric colours do not constitute a conclusive
test of the mechanism responsible for the apparent reddening of Be
stars. Fig.~\ref{bmvvmiplot} shows plots of $B$$-$$V$ and $V$$-$$R$ against
$V$$-$$I$ for Be stars and normal B stars within and around the cluster.  The
$B$$-$$V$ values come from Balona and Jerzykiewicz (\cite{jerky}).  The zero
point of $V$$-$$R$ is arbitrary (since the $R$ magnitude has not been
zero-pointed) but has been adjusted to be close to the value expected for the
normal hot B stars.

In the ($V$$-$$R$, $V$$-$$I$) plot, the Be stars fall on a sequence that is
essentially indistinguishable from the sequence followed by non-H$\alpha$
emitting stars. This overlap is coincidental and is the consequence of the spectral distribution
of the continuum emission from the circumstellar envelope (Kaiser
\cite{kaiser}) within the $R$ and $I$ bands. The position of the Be stars in
the ($B$$-$$V$, $V$$-$$I$) in Fig.~\ref{bmvvmiplot} is suggestive of a
systematic shift to the right of the sequence of normal main-sequence stars,
shown by the continuous line. In this case significant emission due to Paschen
continuum emission from the circumstellar envelope is located in the $I$ band
alone. Hence the Be stars appear shifted to the right relative to the non-Be stars.

Also shown in Fig.~\ref{bmvvmiplot} are the reddening lines from
Taylor(\cite{tay86}).  On the ($B$$-$$V$, $V$$-$$I$) plot, it is clear that the
Be stars do not follow the reddening line, suggesting that the red colours of
the Be stars are not due to circumstellar dust absorption.
 
The effect on $B$$-$$V$ colour of rotational distortion of the atmosphere of 
a Be star has been examined by Collins et al. (\cite{col91}).  
However, even in the case of a star rotating near breakup, the maximum
predicted change in $B$$-$$V$ colour (from the colour of a non-rotating B
star) is only $\sim$0.05 magnitudes (Zorec \& Briot \cite{zor97}), compared to
changes of up to 0.15 magnitudes seen in Fig.~\ref{bmvvmiplot}.  It therefore
seems that this effect is not a major contributor to the reddening of Be stars.

The final possibility is that the red colours of the Be stars compared to the
colours of non-Be stars at the same luminosity (as seen in
Figs.~\ref{fig330cmd}a to \ref{fig2100cmd}a) are due to a real change in the
photospheric flux distribution through the $BVRI$ bands caused by a change in
stellar effective temperature.  However, since the bulk of the flux of these
stars is emitted shortward of $B$, it remains to be seen whether the Be stars
really do have lower effective temperatures than B stars of similar luminosity.
This question will be examined in forthcoming papers which will give spectral
types of the stars, and HST ultraviolet magnitudes from which $T_\mathrm{eff}$
values can be accurately compared between the Be and normal B stars.







\section{Comparison of cluster and field Be star populations}

The Magellanic Clouds offer us an opportunity to examine the field population
of Be stars in different localities in an unbiased way - a task which is
virtually impossible from within our own galaxy due to depth and reddening
effects. For the purposes of our survey we can regard each CCD image as
essentially a volume and magnitude limited sample of the population. 

We have defined the cluster extent by a direct examination of the radial
variation in surface density of bright blue stars ($V$ $<$ 17 and $V$$-$$I$ $<$
0.6), as described in Section~\ref{selection}.  Some stars co-eval with the
cluster stars probably exist outside our adopted radii.  In particular, we note
that Elson et al. (\cite{els87}) have examined the surface density of stars
around a sample of young LMC clusters and suggested that some of these clusters
are tidally unbound with up to 50\% of their total mass in an unbound halo
(although the halo is of very low surface density compared to the background).

\begin{figure}
\resizebox{\hsize}{!}{\includegraphics{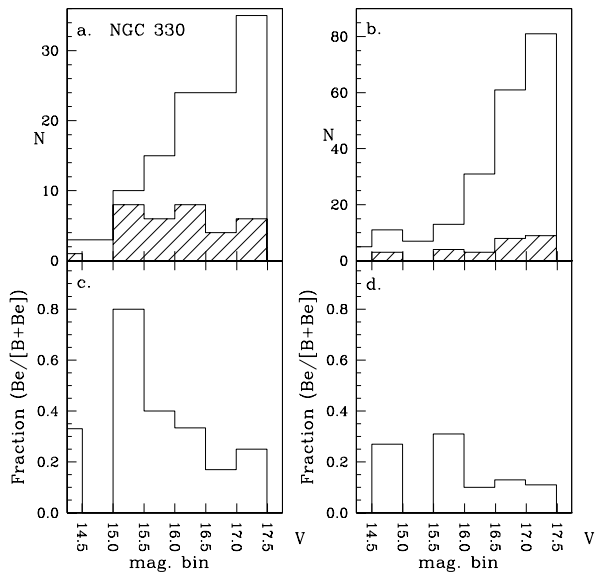}}
\caption{{\bf a.}  Histograms of the number of Be stars (shaded) and
main-sequence stars ($V$$-$$I$ $<$ 0.5, including Be stars) in half-magnitude
bins down the main-sequence of NGC 330. {\bf b.} Same as {\bf a} but for the
field around NGC 330. {\bf c.} The ratio of Be stars to main-sequence stars in
half-magnitude bins down the main-sequence of NGC 330. {\bf d.} Same as {\bf c}
but for the field around NGC 330.}
\label{fig330bef} 
\end{figure}

\begin{figure}
\resizebox{\hsize}{!}{\includegraphics{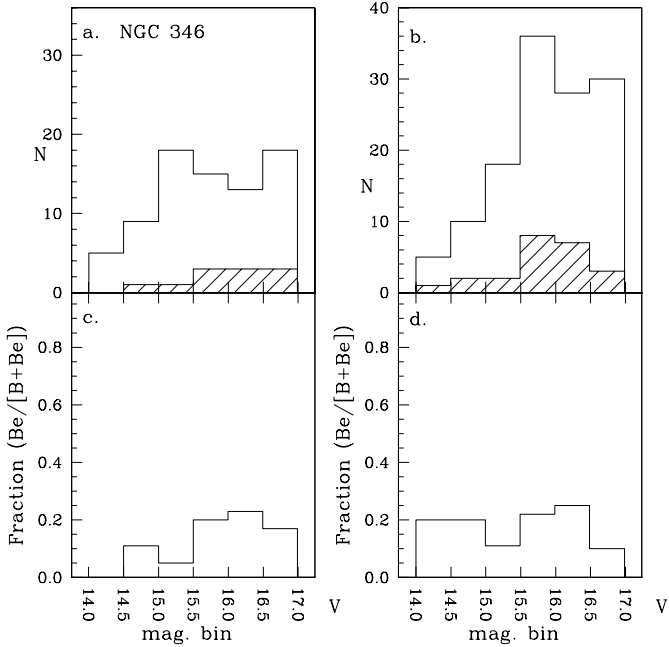}}
\caption{Same as Fig.~\ref{fig330bef} but for NGC 346.}  
\label{fig346bef} 
\end{figure}

\begin{figure}
\resizebox{\hsize}{!}{\includegraphics{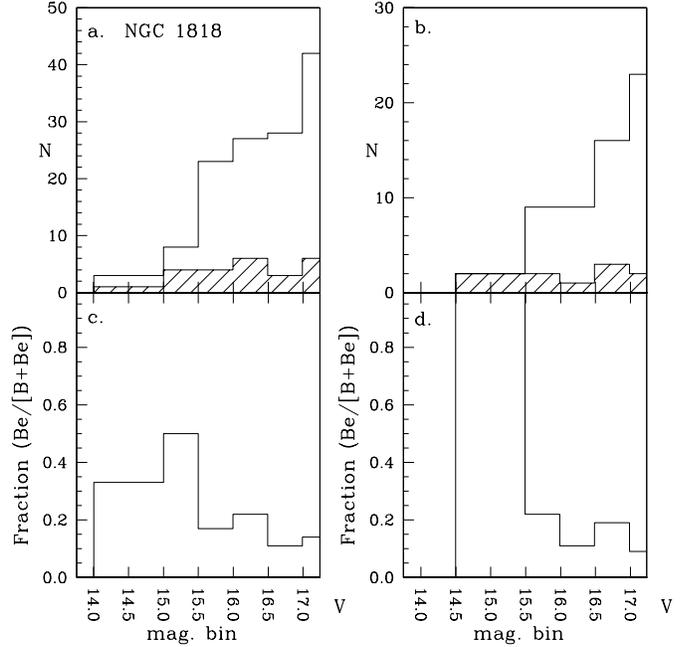}}
\caption{Same as Fig.~\ref{fig330bef} but for NGC 1818.}  
\label{fig1818bef} 
\end{figure}

\begin{figure}
\resizebox{\hsize}{!}{\includegraphics{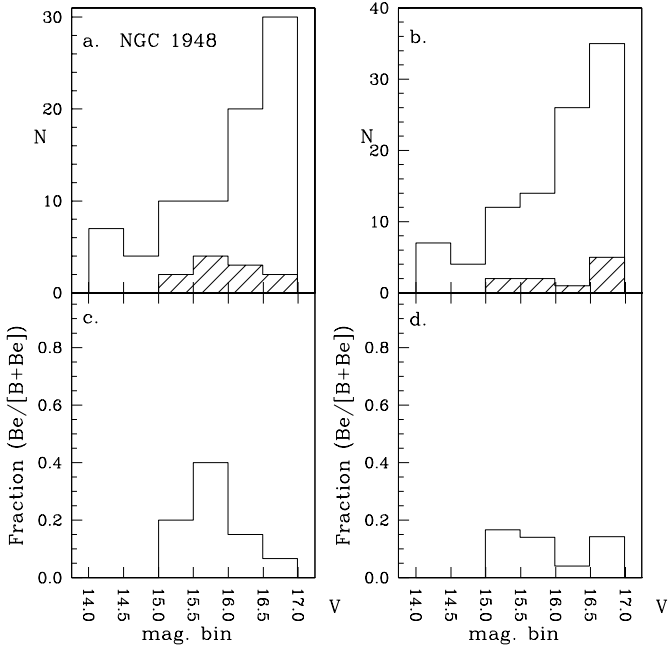}}
\caption{Same as Fig.~\ref{fig330bef} but for NGC 1948.}  
\label{fig1948bef} 
\end{figure}

\begin{figure}
\resizebox{\hsize}{!}{\includegraphics{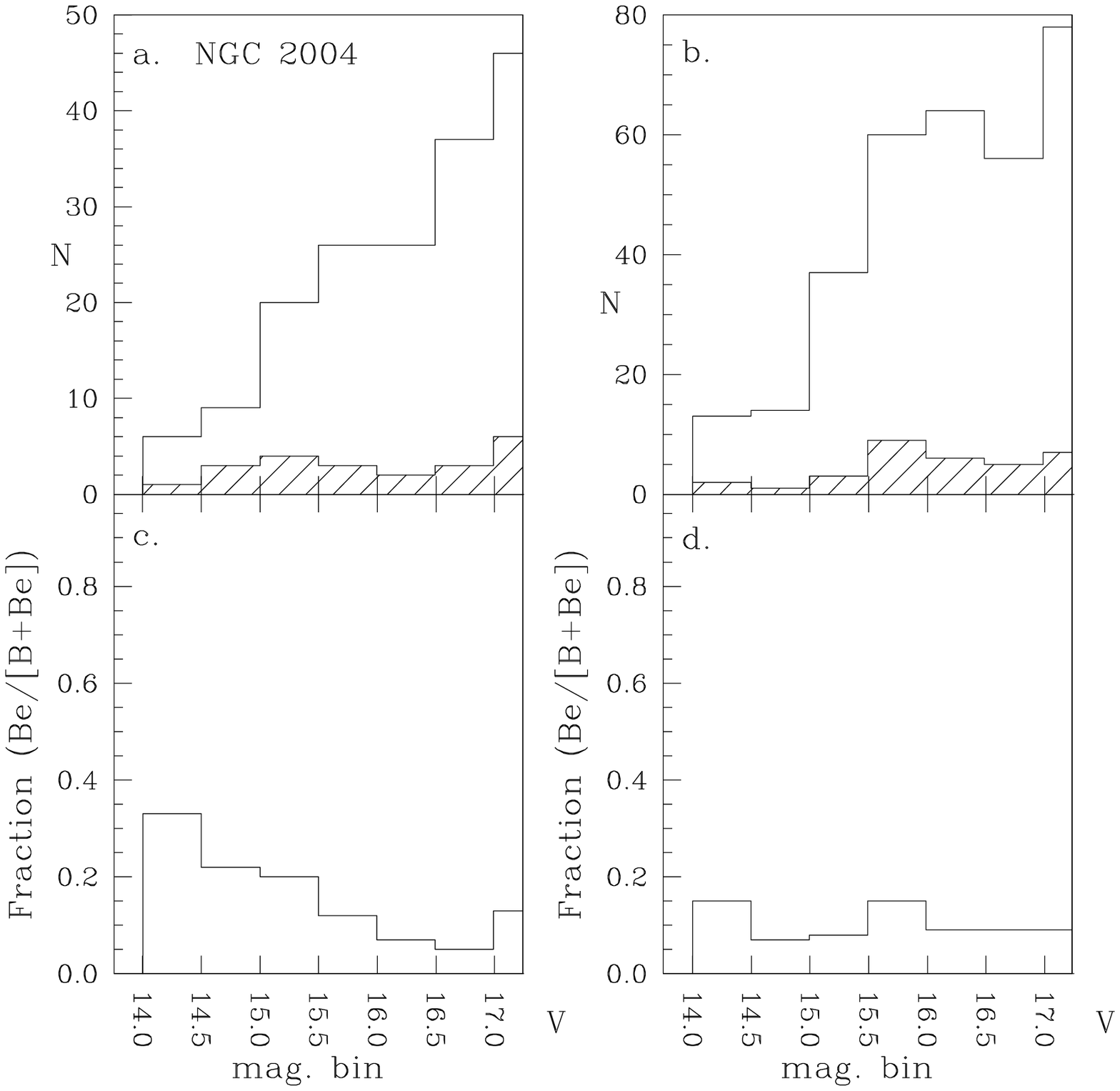}}
\caption{Same as Fig.~\ref{fig330bef} but for NGC 2004.}  
\label{fig2004bef} 
\end{figure}

\begin{figure}
\resizebox{\hsize}{!}{\includegraphics{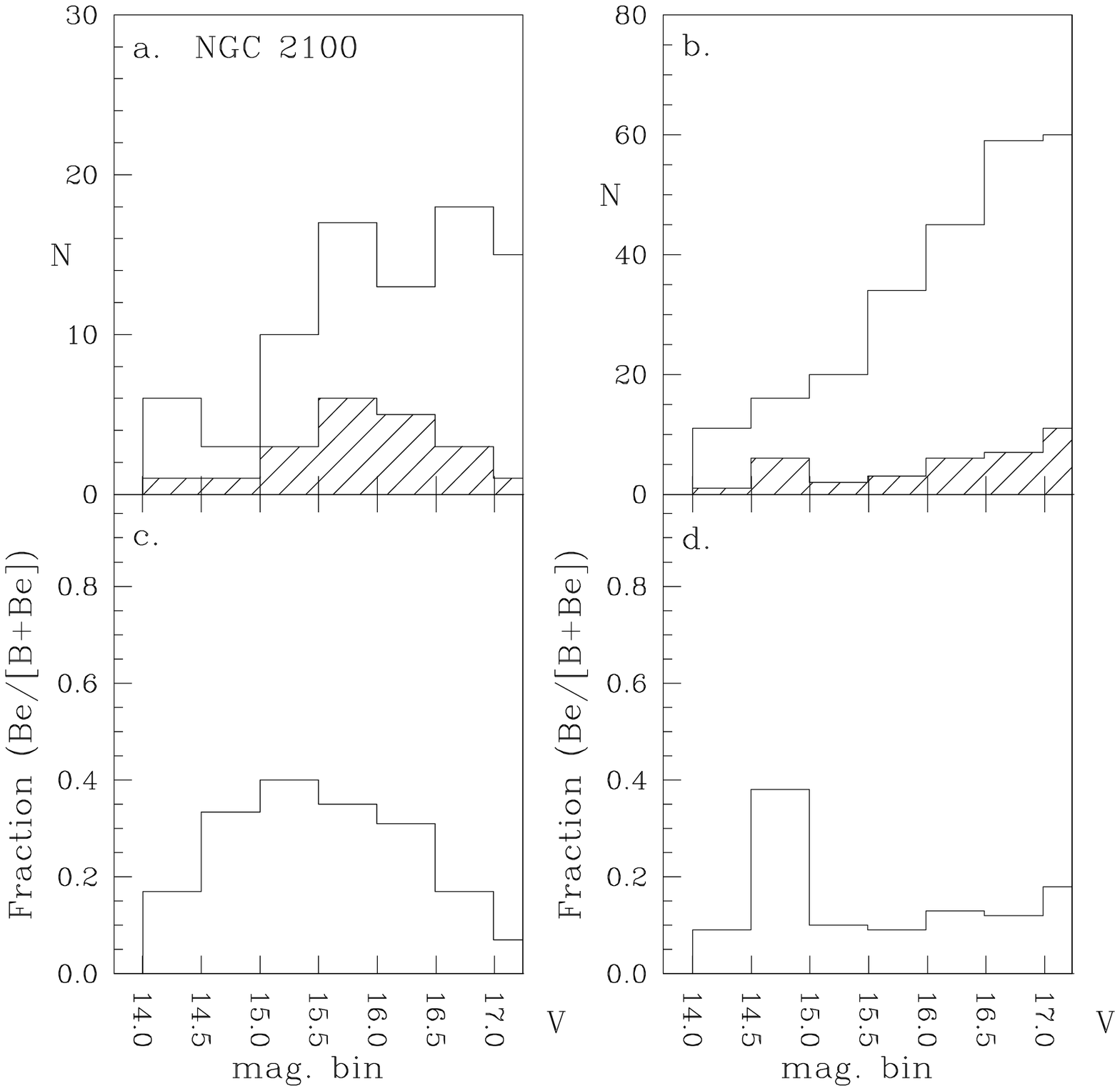}}
\caption{Same as Fig.~\ref{fig330bef} but for NGC 2100.}  
\label{fig2100bef} 
\end{figure}

The CMDs for both the cluster and field of NGC 330, NGC 346, NGC 1948, NGC 2004
and NGC 2100 are very similar (see Figs.~\ref{fig330cmd}b to
\ref{fig2100cmd}b). This indicates that the field populations contain stars of
age similar to that of the cluster.  We have investigated the fraction of stars
along the main-sequence that are Be stars by binning both emission-line and
non-emission-line stars with $V$$-$$I$ $<$ 0.5 into 1/2 magnitude bins in 
$V$.  Figs.~\ref{fig330bef} to \ref{fig2100bef} show our results.  In
Table~\ref{brat}, we give the numbers of Be stars and main-sequence stars with
$V <$ 17 for each cluster and its surrounding field: we also give the ratio of
Be to main-sequence stars (with a 1-$\sigma$ error).

Here we have made the assumption that comparison of the number of Be stars to
B stars within a certain luminosity range is a valid comparison. This
assumption is challenged by the study of Zorec \& Briot (\cite{zor97}) who
claim that the location of Be stars within the CMD is the result of a
visual luminosity excess. The $V$ excess described by Zorec \& Briot rises
from 0.0 at B9 to -0.5 at B0 in a linear manner. Here we do not adjust for $V$
excess as we consider the size of the associated uncertainties in the
correction similar to the size of the correction itself.

The fraction of Be stars in the field populations lies in the range 0.10 to
0.27.  The mean value is a similar to the fraction of Be stars in the Galactic
field which is 0.17 (Zorec and Briot \cite{zor97}).  The fractions of Be stars
in the clusters is quite variable.  The clusters NGC 330 (see also Grebel et
al. \cite{gre92}) and NGC 2100 are particularly rich in Be stars, NGC 1818 has
a moderate richness, while NGC 346, NGC 1948 and NGC 2004 have low Be star
fractions.  In the Galaxy, Zorec and Briot (\cite{zor97}) noted that the Be
star fraction reached a maximum of 0.34 at spectral type B1.  The fraction of
Be stars in NGC 330 (0.34$\pm$0.08) is perfectly consistent with this value.
In a forthcoming paper, the correlation of Be star ratio with spectral type
will be examined.

\begin{table*}
\caption{Be star content of clusters and fields}
\begin{tabular}[]{l|ccccc|ccccc}
\hline
& \multicolumn{5}{c|}{cluster stars}& \multicolumn{5}{c}{field stars}\\
\cline{2-11}
& \multicolumn{3}{c|}{$V$ $<$ 17} & \multicolumn{1}{c|}{16 $<$ $V$ $<$ 17} & $V$ $<$ 16
& \multicolumn{3}{c|}{$V$ $<$ 17} & \multicolumn{1}{c|}{16 $<$ $V$ $<$ 17} & $V$ $<$ 16 \\
\hline
Cluster
& N$_{\rm MS}$ & N$_{\rm Be}$ & \multicolumn{1}{c|}{N$_{\rm Be}$/N$_{\rm MS}$}& 
\multicolumn{1}{c|}{N$_{\rm Be}$/N$_{\rm MS}$} & N$_{\rm Be}$/N$_{\rm MS}$
& N$_{\rm MS}$ & N$_{\rm Be}$ & \multicolumn{1}{c|}{N$_{\rm Be}$/N$_{\rm MS}$}& 
\multicolumn{1}{c|}{N$_{\rm Be}$/N$_{\rm MS}$} & N$_{\rm Be}$/N$_{\rm MS}$\\
\hline
NGC 330 &  79 & 27 & 0.34$\pm$0.08 & 0.25$\pm$0.08 & 0.48$\pm$0.15 & 
 128 & 27 & 0.21$\pm$0.04 & 0.12$\pm$0.04 & 0.19$\pm$0.08 \\
NGC 346 & 78 & 11 & 0.14$\pm$0.05 & 0.19$\pm$0.08 &0.11$\pm$0.05 &
 127 & 23 & 0.18$\pm$0.04 & 0.19$\pm$0.06 & 0.17$\pm$0.05 \\
NGC 1818 & 92 & 19 & 0.21$\pm$0.05 & 0.16$\pm$0.06 & 0.27$\pm$0.10 &
 38 & 10 & 0.27$\pm$0.10 & 0.16$\pm$0.09 & 0.50$\pm$0.2 \\
NGC 1948 & 81 & 11 & 0.14$\pm$0.04 & 0.10$\pm$0.05 & 0.19$\pm$0.08 &
 100 & 10 & 0.10$\pm$0.03 & 0.10$\pm$0.04 & 0.11$\pm$0.06 \\
NGC 2004 & 124 & 16 & 0.13$\pm$0.03 & 0.08$\pm$0.04 & 0.18$\pm$0.06 & 
 224 & 26 & 0.11$\pm$0.02 & 0.09$\pm$0.03 & 0.12$\pm$0.03 \\ 
NGC 2100 & 67 & 19 & 0.28$\pm$0.07 & 0.26$\pm$0.10 & 0.33$\pm$0.11 &
 185 & 25 & 0.14$\pm$0.03 & 0.13$\pm$0.04 & 0.15$\pm$0.05 \\
\hline
\end{tabular}
\label{brat}
\end{table*}

In order to compare the Be star fraction near the main-sequence turnoff with
the Be fraction for less evolved stars, we have divided the stars into two
groups: the brighter stars with $V <$ 16 and the fainter stars with 17 $< V <$
16, all stars having $V$$-$$I$ $<$ 0.6.  The Be star fractions for these groups
are given for the clusters and their surrounding fields in Table~\ref{brat}.
From these ratios, and looking at Figs.~\ref{fig330bef} to \ref{fig2100bef}, it
seems that for most of the clusters and fields there is no statistically
significant difference between the Be star fractions near the main-sequence
turnoff and further down the main-sequence.  There are a few distinct
exceptions to this rule: the clusters NGC 330 and NGC 2004 show a significantly
higher fractions of Be stars near the main-sequence turnoff than at fainter
magnitudes, as does the field of NGC 1818 (where the four brightest
main-sequence stars are all Be stars).

\section{Discussion}

We now examine what our observations imply about the Be mechanism. In the
literature, binarity, pulsation and rotation have been suggested 
as the mechanism for the establishment and maintainence of the
circumstellar material seen in Be stars.

Harmanec and K$\rm\breve{r}$i$\rm\breve{z}$ (\cite{har76}) have suggested that
Be stars comprise a population of interacting binaries. In this model,
elaborated upon by Pols et al. (\cite{pol91}), emission arises from material in
accretion onto the secondary component. The difficulty that remains with this
model is that the expected number of close interacting binaries is too few
to account for the observed population of Be stars (VanBever and Vanbeveren
\cite{van97}). Our observations, particularly the high Be star ratio in NGC
330, do not favour the interacting binary hypothesis as the population of stars
required to be in mass exchanging binary systems near the MS turnoff 
(Abt \cite{abt87}) is too high.

The non-radial pulsator model for Be activity (Vogt and Penrod \cite{vog83};
Baade \cite{baa87}) has gained support from the recent work of Dziembowski et
al. (\cite{dzi93}). In their models, Dziembowski et al. have shown that an
instability domain exists on the main-sequence that almost fills the gap
between $\delta\,$Sct and $\beta\,$Cep variables.  These slowly pulsating
B-type stars span a range of mass from 3 to 6 $\rm{M}_{\odot}$.  The Be star
$\mu\,$Cen has been intensively studied by Rivinius et al. (\cite{riv98}).
From their analysis of photospheric radial velocity variations they have
discerned six discrete periods between 0.27-0.51 days.  Rivinius et al. suggest
that when the majority of these modes are in constructive interference,
expulsion of material and a line emission outburst occurs.

There are problems with associating the Magellanic Cloud Be stars observed here
with the pulsation instability mechanism described above.  Firstly, in
conditions of reduced metallicity such as those within the Magellanic Clouds,
the extent of the pulsation instability domain is reduced (Dziembowski et
al. \cite{dzi93}) and $\beta\,$Cep instability is seemingly extinguished
(Balona \cite{bal92}).  Most importantly, the Be stars observed here are by no means
confined to a range $\la$6 M$_{\odot}$.  Indeed our observations show that the
largest proportion of Be stars occurs at $\sim$14--15 M$_{\odot}$.

The most widely accepted explanation for the Be phenomenon is that it is a
result of the rapid rotation of the B star.  This explanation was originally
put forward by Struve (\cite{str31}). In its modern incarnation, due to
Bjorkman and Cassinelli (\cite{bjo93}), the stellar wind from a rapidly
rotating star is concentrated and confined to a disk around the equatorial
plane of the star. The model of Bjorkman and Cassinelli predicts a maximum
likelihood for disk formation near spectral type B2 (where the stellar angular
velocity required for disk formation is 55\% that of the critical breakup
velocity).  Observations of Galactic Be stars find that the highest Be
star fraction does indeed occur at early spectral types (B1 according to Zorec
and Briot \cite{zor97}). Within our sample spectral types are available for
only five of our flagged Be candidates. We will present further spectral types
in a forthcoming paper.

Assuming the Be phenomenon is related to the formation of an equatorial disk
caused by rapid rotation, we would expect that clusters which form from gas
that has a high angular momentum content would contain a high fraction of Be
stars.  Indeed, the evidence is suggestive that Galactic clusters whose stars
present a higher than average rotational velocity have larger populations of Be
stars (Mermilliod \cite{mermill}). In the Magellanic Clouds, the cluster NGC 330 has a
noticeably ellipsoidal shape which may suggest that the cluster as a whole formed
from material of higher angular momentum and that, as a consequence, the
cluster stars have a rapid rotation rate. This would certainly be
consistent with the high Be star content of NGC 330.  In a future paper we will
examine the rotation velocities of samples of stars within and around the
clusters NGC 330 and NGC 2004 to see if the overall Be star fraction is related
to average cluster star rotation velocity.

The tendency for there to be a higher Be star fraction near the main-sequence
turnoff in some clusters also appears consistent with the rapid rotation
hypothesis for Be star formation.  The evolution of rapidly rotating stars has
been examined by Endal and Sofia (\cite{end79}) and Endal (\cite{end82}). These
models explicitly follow the radial exchange of angular momentum with
evolution.

In the models of Endal and Sofia, stars commencing their main-sequence lives
with relatively slow angular velocity remain slow rotators throughout the whole
of the main-sequence evolutionary phase (although there is a marked increase in
angular velocity of all stars during the core contraction phase associated with
exhaustion of hydrogen in the core, this evolutionary phase is far too short to
account for the observed proportion of Be stars).  Thus slow rotators will
never evolve into Be stars.

In contrast to the slow rotators, the models of Endal and Sofia show that stars
commencing their lives with an angular velocity greater than 56-76\% of the
critical breakup velocity $\omega_{c}$ spin up to the critical velocity over a
moderate fraction of the main-sequence lifetime.  In a cluster where some
fraction of the stars is formed with angular velocity greater than $\sim$50\%
of $\omega_{c}$, those stars nearer the main-sequence turnoff are more likely
to be Be stars since they will have been more spun up by evolution than the
less luminous stars that have not evolved as far through the main-sequence
phase.  This process could explain the concentration of Be stars to the
main-sequence tip seen in NGC 330, NGC 2004 and NGC 1818.  We add the
cautionary note that Endal and Sofia (\cite{end79}) and Endal (\cite{end82})
made a detailed study of only a single star which had a mass of 5 M$_{\odot}$,
much less than the turnoff masses of $\sim$15 M$_{\odot}$ which is appropriate
for the clusters observed here.  For comparison with Magellanic Cloud data,
models such as those of Endal and Sofia need to be made at higher masses and
lower metallicities.

Other factors which might influence the Be star fraction are metallicity and
age.  However, neither of these factors seem to influence the Be star fractions
in our data.  To examine abundance dependence, we note that the two SMC
clusters are probably more metal poor than the LMC clusters by a factor of
about two (Russell and Bessell \cite{rus89}).  However, in the SMC, NGC 330 has
a high Be star content while NGC 346 does not.  At the higher abundance of
the LMC, clusters with a range of Be star content are present, and the range
is similar to the range in the SMC in spite of the abundance difference 
between SMC and LMC.  Turning to age,
we note that the clusters studied here have ages of $\sim$1--3$\times$10$^{7}$
years, although these ages are uncertain by factors of 2--3 because of
uncertainties in the amount of convective core overshoot which occurs in real
stars.  For comparison of Be star fractions with age, we take from the
literature the following ages: NGC 330, 20 Myr (Chiosi et al. \cite{chi95});
NGC 346, 13 Myr (Massey et al. \cite{mas89}); NGC 1818, 30 Myr (Will et
al. \cite{will2}); NGC 1948, 15$\pm$5 Myr (Vallenari et al. \cite{val93}); NGC
2004, 20 Myr (Caloi and Cassatella \cite{cal95}); NGC 2100, 15 Myr (Cassatella
et al. \cite{cas96}).  There does not appear to be any correlation of Be star
fraction with age: the cluster NGC 330 with the highest Be star fraction has an
intermediate age and the oldest cluster (NGC 1818) has an intermediate Be star
fraction.  This result contrasts with the results of Mermilliod
(\cite{mermill}) who noted a trend of decreasing Be fraction with increasing
age amongst young Galactic clusters.

\section{Summary} 

We have presented the results of a photometric survey for Be stars in six
fields centred on NGC 330 and NGC 346 in the SMC, and NGC 1818, NGC 1948, NGC
2004 and NGC 2100 in the LMC. In all six of these young populous clusters we
find large numbers of Be stars.

The Be stars are shown to be redder in $V$$-$$I$ than main-sequence stars of
similar $M_{\rm V}$. Furthermore, the displacement to the red increases with
increasing H$\alpha$ emission line strength.  Some suggested causes for the
redness of Be stars were examined using the widely accepted model that Be
stars are main-sequence stars with a surrounding disk. Our photometric
observations suggest that the redness of the stars is due to emission from the
circumstellar envelope, in line with the results of detailed spectrophotometric
studies based upon galactic samples.

The fraction of main-sequence stars that are Be stars varies significantly
between the various clusters from $\sim$0.10--0.34.  The average value is close
to the average fraction of $\sim$0.17 seen in the Galaxy, and the maximum value
is also similar to the maximum Galactic value, which occurs at spectral type
B1. It was noted that the cluster with the maximum Be star fraction, NGC 330,
is distinctly elliptical, suggesting that the stars in it formed out of rapidly
rotating material and that there is a connection between rapid rotation and the
Be phenomenon.  No connection between Be star fraction and age or metallicity
was evident.  In some clusters, the Be stars tend to be concentrated toward the
main-sequence turnoff.  This is consistent with evolutionary models which
predict that stars which form with rotation velocities more than $\sim$50\% of
the breakup value spin up as they evolve, reaching the breakup velocity over a
moderate fraction of the main-sequence lifetime.

\begin{acknowledgements} 
We thank Alistair Walker for making available his
photometry prior to publication. SCK acknowledges the support of an Australian
Postgraduate Award scholarship. We thank our referee Eva Grebel for her useful
comments.
\end{acknowledgements}

\newpage

\begin{figure*}
\resizebox{\hsize}{!}{\includegraphics*[74,197][574,716]{ds7804f14.eps}}
\caption{The finding chart ($R$ image) for the Be stars around NGC 330.  
The full field is 10$\arcmin\times$10$\arcmin$ while the
inner rectangle is $7\farcm3\times7\farcm9$ and is the part of the field on
which the search for stars with H$\alpha$ was performed.  North is to the right
and east is at the top. }  
\label{fig330field} 
\end{figure*}

\begin{figure*}
\resizebox{\hsize}{!}{\includegraphics*[74,197][574,716]{ds7804f15.eps}}
\caption{Same as Fig.~\ref{fig330field} but for stars near the
core of NGC 330.}  
\label{fig330cluster} 
\end{figure*}

\begin{figure*}
\resizebox{\hsize}{!}{\includegraphics*[74,197][574,716]{ds7804f16.eps}}
\caption{The finding chart for the Be stars around NGC 346.
The field size is 10$\arcmin$ square and the orientation is north to
the right, east at the top. } 
\label{fig346field}
\end{figure*}

\begin{figure*}
\resizebox{\hsize}{!}{\includegraphics*[74,197][574,716]{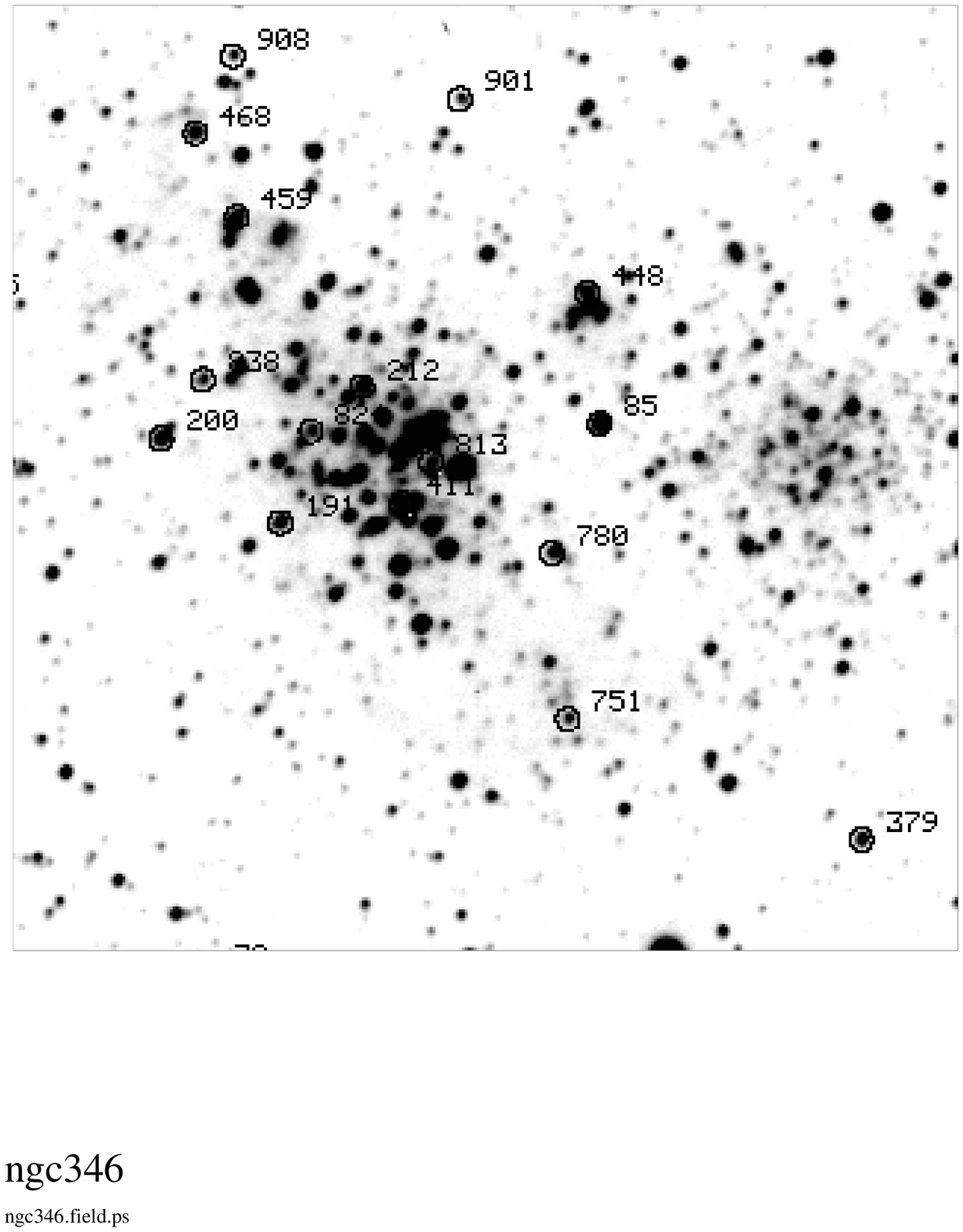}}
\caption{Same as Fig.~\ref{fig346field} but for stars near the 
core of NGC 346.}  
\label{fig346cluster} 
\end{figure*}

\begin{figure*}
\resizebox{\hsize}{!}{\includegraphics*[74,197][574,716]{ds7804f18.eps}}
\caption{The finding chart for the Be stars around NGC 1818. 
The full field is 10$\arcmin\times$10$\arcmin$ whilst the inner
rectangle in which the search for stars with H$\alpha$ emission was performed
is $7\farcm6\times9\farcm2$.  North is to the right and east is at the top.}  
\label{fig1818field} 
\end{figure*}

\begin{figure*}
\resizebox{\hsize}{!}{\includegraphics*[74,197][574,716]{ds7804f19.eps}}
\caption{Same as Fig.~\ref{fig1818field} but for stars near the
core of NGC 1818.}  
\label{fig1818cluster} 
\end{figure*}

\begin{figure*}
\resizebox{\hsize}{!}{\includegraphics*[74,197][574,716]{ds7804f20.eps}}
\caption{The finding chart for the Be stars around NGC 1948.
The full field is 10$\arcmin\times$10$\arcmin$ whilst the inner
rectangle in which the search for stars with H$\alpha$ emission was performed
is $7\farcm6\times9\farcm2$.  North is to the right and east is at the top.}
\label{fig1948field} 
\end{figure*}

\begin{figure*}
\resizebox{\hsize}{!}{\includegraphics*[74,197][574,716]{ds7804f21.eps}}
\caption{The finding chart for the Be stars around NGC 2004.
The field size is 10$\arcmin\times$10$\arcmin$.  
North is to the right and east is at the top.}
\label{fig2004field} 
\end{figure*}

\begin{figure*}
\resizebox{\hsize}{!}{\includegraphics*[74,197][574,716]{ds7804f22.eps}}
\caption{Same as Fig.~\ref{fig2004field} but for stars near the
core of NGC 2004.}  
\label{fig2004cluster} 
\end{figure*}

\begin{figure*}
\resizebox{\hsize}{!}{\includegraphics*[75,197][574,716]{ds7804f23.eps}}
\caption{The finding chart for the Be stars around NGC 2100.
The field size is 10$\arcmin\times$10$\arcmin$.
North is to the right and east is at the top.}  
\label{fig2100field} 
\end{figure*}

\begin{figure*}
\resizebox{\hsize}{!}{\includegraphics*[74,197][574,716]{ds7804f24.eps}}
\caption{Same as Fig.~\ref{fig2100field} but for stars near the 
core of NGC 2100.}
\label{fig2100cluster} 
\end{figure*}

\end{document}